\documentclass[12pt,a4paper]{article}%
\usepackage{epsfig}
\usepackage{amsmath}
\usepackage{amsfonts}
\usepackage{amssymb}
\usepackage{graphicx}
\usepackage[T1]{fontenc}
\usepackage[utf8]{inputenc}
\usepackage{cite}
\usepackage{authblk}%
\providecommand{\U}[1]{\protect\rule{.1in}{.1in}}
\begin{document}

\title{ On the $\eta$ pseudo $\mathcal{PT}$ symmetry theory for non-Hermitian
Hamiltonians: time-dependent systems}
\author{Mustapha Maamache\thanks{E-mail: maamache@univ-setif.dz}\\Laboratoire de Physique Quantique et Syst\`{e}mes Dynamiques,\\Facult\'{e} des Sciences, Universit\'{e} Ferhat Abbas S\'{e}tif 1, S\'{e}tif
19000, Algeria. }
\date{}
\maketitle

\begin{abstract}
\ In the context of non-Hermitian quantum mechanics, many systems are known to
possess\ a pseudo $\mathcal{PT}$ symmetry , i.e. the non-Hermitian Hamiltonian
$H$ is related to its adjoint $H^{\dagger}$ via the relation, $H^{\dagger
}=\mathcal{PT}H\mathcal{PT}$ . We propose a derivation of pseudo
$\mathcal{PT}$ symmetry and $\eta$ -pseudo-Hermiticity simultaneously for the
time dependent non-Hermitian Hamiltonians by intoducing a new symmetry
operator $\tilde{\eta}(t)=\mathcal{PT}\eta(t)$ that not satisfy the
time-dependent quasi-Hermiticity relation but obeys the Heisenberg evolution
equation. Here, we solve the $SU(1,1)$ time-dependent non-Hermitian
Hamiltonian and we construct a time-dependent solutions by employing this new
symmetry operator and discuss a concrete physical applications of our results.

Dedicated to the memory of my mother \textbf{Djabou Zoulikha} and to
\textbf{Djabi Smail}

\end{abstract}

\section{Introduction}

Quantum mechanics (QM) forms the basis of modern physics, developed in the
early 20th century, it has introduced a radical transformation in our
understanding of the fundamental nature of reality. Providing a framework for
describing matter at the smallest scales, quantum mechanics emerged in
response to the limitations of classical physics. Quantum mechanics lies the
concept of the wave function, the wave function is a mathematical
representation denoted by $\left\vert \psi\right\rangle $ defined in the
Hilbert space. The wave function evolves over time according to
Schr\"{o}dinger's equation, a fundamental equation in quantum mechanics%
\begin{equation}
i\frac{\partial\left\vert \psi(t)\right\rangle }{\partial t}=H\left\vert
\psi(t)\right\rangle \label{SE}%
\end{equation}
where the Hamiltonian $H$ represents the total energy of a system. The theory
was built upon a set of fundamental postulates, one of these postulates
involve the hermiticity of the Hamiltonian $H$ which represents the total
energy of a system. Consequently, the Hermiticity of the Hamiltonian
guarantees a real energy-spectrum and the probability invariance during the
time evolution.

However, In 1998, Bender et al \cite{Bender1,Bender4,Bender8', Bender8"}
claimed that $\mathcal{PT}$-symmetry is a more general form of quantum
mechanics than Hermitian quantum mechanics. They were able to identify a
certain class of Non-Hermitian (NH) Hamiltonians\ that admits discrete real
spectrum and they attributed the reality of this spectrum to the unbroken
symmetry under combined parity ($\mathcal{P}$) and time reversal
($\mathcal{T}$) transformations, this led to the development of a new theory
known as $\mathcal{PT}$-symmetry theory(i.e$.$\textrm{ }$H$ commutes
with\textrm{ }$\mathcal{PT}$).The parity operator$\mathcal{P}$\textrm{ }has
the following effect;\textrm{ }$\mathcal{P}$ :$\left\{  x\rightarrow-x\text{,
\ }p\rightarrow-p\right\}  $ \ and the effect of the time-reversal is
$\mathcal{T}:\left\{  x\rightarrow x\text{ , }p\rightarrow-p\text{ ,
}i\rightarrow-i\right\}  $.

They asserted that: i) the\textrm{ }$\mathcal{PT}$-symmetric and the
$\mathcal{CPT}$-symmetric together form a physically reasonable alternative to
Hermitian quantum mechanics. ii)\textrm{ }$\mathcal{PT}$\textrm{ }-symmetric
Hamiltonians can be used to define real energy eigenvalues for a quantum
system. In addition, unitary time evolution can be defined if the
$\mathcal{CPT}$\textrm{ - }inner\textrm{ }product is used to define a Hilbert space.

The inner-product of the $\mathcal{PT}$-symmetric Hamiltonian's eigenfunctions
$\left\vert \psi_{n}\right\rangle $ and $\left\vert \psi_{m}\right\rangle $ is
not positive definite
\begin{equation}
\left\langle \psi_{n},\psi_{m}\right\rangle _{\mathcal{PT}}=\int dx\left[
\mathcal{PT}\text{ }\psi_{n}(x)\right]  \psi_{m}(x)=(-1)^{n}\delta_{nm},
\label{ps2}%
\end{equation}
\textrm{ }and this is problematic as the inner-product is traditionally viewed
as a probability in Hermitian quantum mechanics. To overcome the issue of
negative norm, Bender et al \cite{Bender4}\textbf{ }introduced the
linear\textbf{ }operator $\mathcal{C}$, with eigenvalue\textbf{s} $\pm1$:
$\mathcal{C}%
{{}^2}%
=1$. This operator commutes with the $\mathcal{PT\ }$operator but not with
the\textbf{ }$\mathcal{P}$\ and $\mathcal{T}$\ operators separately\textbf{.
}They define a new structure of inner product, known as the $\mathcal{CPT\ }%
$\ inner product%

\begin{equation}
\left\langle f,g\right\rangle _{\mathcal{CPT}}\emph{=}\int\emph{dx}\left[
\mathcal{CPT}\text{ }f(x)\right]  \emph{g(x).} \label{ps3}%
\end{equation}

Like the $\mathcal{PT}$\ inner product (\ref{ps2}), this inner product is also
phase independent and conserved in time. The inner product (\ref{ps3}) is
positive definite because (the operator) $\mathcal{C}$\ contributes $-1$\ when
it acts on states with negative $\mathcal{PT}$\ norm. This new operator
$\mathcal{C}$\ resembles the charge-conjugation operator in quantum field
theory. However, the precise meaning of $\mathcal{C}\ $is that it represents
the measurement of the sign of the $\mathcal{PT}$\ norm in (\ref{ps2}) of an
eigenstate. Specifically\textbf{\ }%

\begin{equation}
\mathcal{C}\psi_{n}(x)=(-1)^{n}\psi_{n}(x). \label{ps4}%
\end{equation}

And the problem of negative norms has been fixed \ by the introduction of the
new operator\textrm{ }$\mathcal{C},$\textrm{ }it was found that a new
dynamically determined inner-product could be defined
\begin{equation}
\left\langle \psi_{n},\psi_{m}\right\rangle _{\mathcal{CPT}}=\int dx\left[
\mathcal{CPT}\text{ }\psi_{n}(x)\right]  \psi_{m}(x)=\delta_{nm},
\end{equation}
indeed,the issue of negative norm is effectively resolved.

Mostafazadeh \cite{Mostafa1,Mostafa7,Mostafa2,Mostafa3} pointed out that the
condition that a Hamiltonian $H$ \ be $\mathcal{PT}$-symmetric can be
understood more generally as a special case of pseudo-Hermiticity. An operator
$H$ is said to be pseudo-Hermitian \cite{scholtz} if
\begin{equation}
H^{\dagger}=\eta H\eta^{-1}, \label{1'}%
\end{equation}
where $H^{+}$ is the conjugate of $H$ and $\eta=\rho^{+}\rho$\ is a linear
Hermitian operator called the metric operator. \ These systems share the same
real energy eigenvalues associated to the Hermitian Hamiltonian $h=\rho
H\rho^{-1}\ $via\ the Dyson map $\rho$. The system does not generally lead to
unitary time evolution. However, like the case of $\mathcal{PT}$-symmetric
non-Hermitian systems, presence of the additional operator $\eta$ in the
pseudo-Hermitian theories allows one to de ne a new inner product in the
fashion
\begin{equation}
\left\langle \psi_{n},\psi_{m}\right\rangle _{\eta}=\left\langle \psi
_{n}\right\vert \eta\left\vert \psi_{m}\right\rangle =\delta_{nm}, \label{IP'}%
\end{equation}

Later a novel concept of the pseudo parity-time (pseudo\textbf{-}%
$\mathcal{PT}$\textbf{\ ) }symmetry, which can be understood as a case of
pseudo-Hermiticity, was introduced in \cite{maamache2} in order to connect the
non-Hermitian Hamiltonian\textbf{\ }$H$\textbf{\ }to its Hermitian
conjugate\textbf{\ }$H^{\dagger}$ . An operator $H$ is said to be pseudo
$\mathcal{PT}$\textbf{\ }-symmetric if%
\begin{equation}
H^{\dag}\mathcal{=PT}H\mathcal{PT}, \label{H+}%
\end{equation}
where in the expression of the inner product (\ref{IP'}), the
metric\textbf{\ }$\eta$\textbf{\ \ }is replaced by $\mathcal{PT}.$ It has been
observed that certain systems do not exhibit exact $\mathcal{PT}$-symmetries,
but they can manifest a distinct form of pseudo $\mathcal{PT}$-symmetry, that
extends beyond the traditional $\mathcal{PT}$-symmetry. \thinspace These
systems, characterized by non-self-adjoint Hamiltonians $H^{\dag}$, deviate
from the standard Hermitian framework and give rise to complex eigenvalues and
non-unitary time evolution, Similar to $\mathcal{PT}$-symmetry, pseudo
$\mathcal{PT}$-symmetry is characterized by the Eq.(\ref{H+}).

The concept of pseudo $\mathcal{PT}$-symmetry has found wide application in
various domains, including periodically high-frequency driven systems
\cite{xia1}, time periodic non-hermitian Hamiltonian systems \cite{maamache3},
optical systems \cite{xia2}, and even the Dirac equation \cite{maamache4}.

Aforementioned, we have discussed pseudo Hermitian and pseudo $\mathcal{PT}%
$-symmetric Hamiltonians, and argued that they constitute independent classes
relating the adjoint $H^{\dag}$ to $H$.

Despite the prolific literature on $\mathcal{PT}$-symmetric quantum mechanics
\cite{Bender1,Bender4,Bender8', Bender8", 1, 2, 3, 4, 5, 6}, where the
spectrum of $\mathcal{PT}$-symmetric Hamiltonians and the properties of their
eigenstates has been analyzed in detail \cite{2}, the pseudo $\mathcal{PT}$
-symmetric Hamiltonians are generally overlooked. In the next section, we
recall the fundamental principles of the time independent $\eta$
pseudo$\mathcal{PT}$-symmetry theory where the new symmetry operator
$\tilde{\eta}$, obtained by multiplying the $\mathcal{PT}$ operator by $\eta$,
commutes with the non-Hermitian Hamiltonian and thus makes the $\eta$ pseudo
$\mathcal{PT}$ symmetric regime physically significant and gives our system
the same properties as in the case of symmetric $\mathcal{PT}$ systems.
\ Furthermore, we generalize this framework to time-dependent systems.

In order to generalize this framework to time-dependent systems, we create an
elegant framework for metric transformations of time-dependent non-Hermitian
Hamiltonians. To make sense of these systems we need to calculate the symmetry
operator $\tilde{\eta}(t)$ and show that obey the Heisenberg equation and not
the time-dependent quasi-Hermiticity relation as in the pseudo Hermitian case
\cite{7', 8-}. At the end of this paper, we outlined an application in time
dependent $SU(1,1)$ system. Finally, a conclusion finalize this work.

\section{\bigskip$\eta$ pseudo $\mathcal{PT}$ symmetry theory}

In order to introduce the theory of $\eta$ pseudo $\mathcal{PT}$-symmetry ,
let us substitute the expression of $H^{\dag}$ (\ref{1'}) into (\ref{H+})
\begin{equation}
H=\mathcal{PT}\eta H\eta^{-1}\mathcal{PT}=\tilde{\eta}H\tilde{\eta}^{-1},
\label{NN}%
\end{equation}
this allows us to define a symmetry operator $\tilde{\eta}$ such that%

\begin{equation}
\tilde{\eta}=\mathcal{PT}\eta\text{\ },\text{ } \label{00}%
\end{equation}
the result (\ref{NN}) indicated the possibility to compensate the broken
symmetry of a Hamiltonian by the presence of $\tilde{\eta}$-symmetry. The
spectra of many non-hermitian Hamiltonian $H$ are indeed real if they are
invariant under the action of symmetry operator $\tilde{\eta}$, i.e, $\left[
H,\tilde{\eta}\right]  =0$ and if the energy eigenstates are invariant under
the operator $\tilde{\eta}$.

The main result of the paper is the introduction of the operator $\tilde{\eta
}$ in (\ref{NN}), that results as a combination of equations (\ref{1'}) and
(\ref{H+}). This operator cannot serve as a metric operator, since $\eta$ is a
positive metric and $\mathcal{PT}$ not positive the product is also not
positive, it is obviously a symmetry operator. 

Therefore, the equation (\ref{00}) can be reformulated as follows%

\begin{equation}
\tilde{\eta}=\mathcal{PT}\widetilde{\rho}^{^{\dag}}\widetilde{\rho
}.\label{nor}%
\end{equation}
Which allows us to define an invertible operator $\widetilde{\rho}$ (Dyson
operator) such that%

\begin{equation}
\widetilde{\rho}=\mathcal{PT}\rho\text{\ },\text{ }%
\end{equation}
the result (\ref{NN}) indicated that in presence of the $\tilde{\eta}%
$-symmetry, the possibility to obtain the unbroken symmetry. The operator
$\tilde{\eta}$ possesses the following properties: i) $\tilde{\eta}%
=\tilde{\eta}^{-1}$, ii) $\left[  H,\tilde{\eta}\right]  =0$, iii)
$\tilde{\eta}\neq\tilde{\eta}^{+}$, iv) $\eta$ $=\mathcal{PT}\eta
^{-1}\mathcal{TP}$ , for more details see Ref.\cite{Absi}.

The parameter space over which the spectrum of the $\tilde{\eta}$ -symmetric
Hamiltonian is real is called the region of unbroken $\tilde{\eta}$ symmetry,
in which case every eigenfunction of the $\tilde{\eta}$ -symmetric Hamiltonian
is also an eigenfunction of the $\tilde{\eta}$ operator. Broken $\tilde{\eta}$
symmetry occurs when some eigenvalues of the $\tilde{\eta}$ -symmetric
Hamiltonian become complex; thus, the Hamiltonian and the $\tilde{\eta}$
operator no longer share simultaneous eigenfunctions.

Another possibility to explain the reality of the spectrum is to relate the
non-Hermitian Hamiltonian, $H\neq H^{\dag}$, to a Hermitian Hamiltonian, $h$
$=$ $h^{+}$, through the action of a map $\rho$ such that%

\begin{equation}
H=\mathcal{PT}\rho^{\dag}h\rho^{-1\dag}\mathcal{PT}, \label{hh}%
\end{equation}
where the conjugate of the Dyson operator $\rho$ satisfies $\rho^{\dag
}=\mathcal{PT}\rho^{-1}\mathcal{PT}$ and the time-independent metric is
$\eta=\rho^{\dag}\rho$ .

At first, the developments of $\mathcal{PT}$ and non-Hermitian pseudo quantum
mechanics were mainly concerned with the study of time-independent systems,
afterwards the study of time-dependent systems has been addressed in different
manners \cite{5', 6', 7', 8-, 9', 10, 11, Koussa2, 12, Da1, Da2, Ni}. The
contributions \cite{7', 8-} have advanced the grounds for treating
time-dependent non-Hermitian Hamiltonians through time dependent Dyson maps
and metric operator. It has been demonstrated that the time-dependent
quasi-Hermiticity relation
\begin{equation}
i\hbar\dot{\eta}\left(  t\right)  =H^{\dag}\left(  t\right)  \eta\left(
t\right)  -\eta\left(  t\right)  H\left(  t\right)  , \label{PHH1}%
\end{equation}
can be solved consistently in such a scenario for a time-dependent Dyson map
and time-dependent metric operator, respectively. In other words, $H(t)$ being
non-Hermitian has to be allied to its Hermitian conjugate $H^{^{\dag}}(t)$
through the relation (\ref{PHH1}). In light of the above discussion, one
important question motivates our work here: How can we treat a non-Hermitian
time-dependent $\eta$ pseudo $\mathcal{PT}$ -symmetry and investigate the possibility\ of\ finding\ the\ exact\ solution\ of\ the\ Schr\"{o}dinger\ equation?\ 

Before\ getting\ to\ the\ heart\ of\ the\ matter,\ we\ will\ recall\ that\ when\ time-reversal\ was\ introduced\ into\ quantum\ mechanics\ by\ Wigner,\ the\ time\ reflection\ of\ the\ $i\partial
/\partial t$\ operator\ was\ achieved\ not\ by\ replacing\ $t$\ by\ $-t$%
\ but\ rather\ by\ taking$\mathcal{\ T}$%
\ \ to\ be\ an\ antiunitary\ operator\ that\ transforms\ $i$\ into\ $-i$%
,\ time\ $t$%
\ was\ treated\ as\ a\ c-number\ parameter\ that\ is\ not\ affected\ by
$\mathcal{T}$ \ \cite{CM, Mannheim, Roberts}
\begin{equation}
\mathcal{T}:\text{ }\mathcal{\ }p\mathcal{\rightarrow}-p\text{ ,
\ }x\rightarrow x\text{\ \ and \ }i\rightarrow-i.
\end{equation}
Although some authors \cite{Moiseyev2011,Luo2013,Lian2014,Gu1} use the fact
that the action of $T$ on time changes $t$ to $-t$.

In order to introduce the time-dependent symmetry operator $\tilde{\eta
}(t)=\mathcal{PT}\eta(t)$ which will interpreted as $\eta(t)$-pseudo
$\mathcal{PT}$ \textbf{operator}.\ Let us embark from the quasi-hermiticity
relation (\ref{PHH1}) proposed in \cite{7', 8-}. In the time dependent case,
we keep $\mathcal{T}$ and $\mathcal{P}$ as time-independent, we may then
demonstrate that the relation (\ref{Heisenberg}) can be obtained by using the
equation (\ref{PHH1}) and differentiating the $\tilde{\eta}(t)$\textbf{ }with
respect to\textbf{ }$t$. In this way we compute\textbf{ }$\partial_{t}%
(\tilde{\eta})=\mathcal{PT}$ $\partial_{t}\eta\left(  t\right)  =\mathcal{PT}%
(-iH^{\dag}\left(  t\right)  \eta\left(  t\right)  +i\eta\left(  t\right)
H\left(  t\right)  ).$ It then follows immediately
\begin{equation}
\partial_{t}(\tilde{\eta})=i\left[  H,\tilde{\eta}\right]  .\label{Heisenberg}%
\end{equation}

This last equation constitute a new result and indicates that the new symmetry
operator $\tilde{\eta}(t)$ obeys the Heisenberg equation contrary to Eq.
(\ref{H+}) defined by Fring and Moussa \cite{7', 8-}. This implies that
$\tilde{\eta}(t)$ evolves in a manner consistent with the dynamics of the
system described by the Hamiltonian $H$. \ We refer to Eq. (\ref{Heisenberg})
as the time-dependent quasi-Hermiticity relation in the $\eta$ -pseudo
$\mathcal{PT}$ \ theory. Let us see this in detail for an example by solving
(\ref{Heisenberg}).

\section{Application: Non Hermitian time-dependent $SU(1,1)$ systems}

\subsection{\bigskip Solutions of the $\eta$ pseudo $\mathcal{PT}$
\ quasi-Hermiticity relation}

The system we wish to investigate here is related to the time dependent pseudo
$\mathcal{PT-}$symmetric Hamiltonian written in terms of the $SU(1.1)$
algebra
\begin{equation}
H(t)=\Omega(t)\hat{K}_{0}+iG(t)\left(  \hat{K}_{+}+\hat{K}_{-}\right)  ,
\label{0}%
\end{equation}
where the real \ parameter\textbf{ }$\Omega(t)$ and $G(t)$
are\ not\ affected\ by $\mathcal{T}$ ($\Omega$ being the time dependent
driving frequency and $G$\textbf{ }a time dependent coupling parameter).$\ $%
The Hamiltonian $H(t)$ is not Hermitian.

The generators of the corresponding $SU(1,1)$ Lie algebra are defined by the
following commutation relations
\begin{align}
\left[  \hat{K}_{0},\hat{K}_{\pm}\right]   &  =\pm\hat{K}_{\pm}\,,\nonumber\\
\left[  \hat{K}_{+},\hat{K}_{-}\right]   &  =-2\hat{K}_{0}, \label{commute}%
\end{align}
$\hat{K}_{0}$ is hermitian while $\hat{K}_{+}=\left(  \hat{K}_{-}\right)
^{^{\dag}}$ and $\ \hat{K}_{-}=\left(  \hat{K}_{+}\right)  ^{^{\dag}}$.

\bigskip And the $SU(1,1)$ Lie algebra has a realization in terms of boson
creation and annihilation operators $\overset{\symbol{94}}{a}^{\dag}$and
$\overset{\symbol{94}}{a}$ \textbf{\ }such that $\hat{K}_{0}=\frac{1}%
{2}(\overset{\symbol{94}}{a}^{\dag}\overset{\symbol{94}}{a}+\frac{1}{2})$ ,
\ $\hat{K}_{+}=\frac{1}{2}(\overset{\symbol{94}}{a}^{\dag})^{2},\hat{K}%
_{-}=\frac{1}{2}(\overset{\symbol{94}}{a})^{2}$. As a matter of fact the
Hamiltonian $H(t)$ (\ref{0}) describes a non Hermitian harmonic-oscillator
$H(t)=\left(  \Omega-2iG\right)  p^{2}+\left(  \Omega-2iG\right)  q^{2}$ in
the coordinate and momentum representation of boson operators $\overset
{\symbol{94}}{a}^{\dag}=(q-ip)/\sqrt{2}$, \ $\overset{\symbol{94}}%
{a}=(q+ip)/\sqrt{2}$.

Under the $\mathcal{PT}$ transformation the $SU(1,1)$ generators transform as
\begin{equation}
\mathcal{PT}\hat{K}_{0}\mathcal{PT}\mathcal{=}\hat{K}_{0},\mathcal{PT}\hat
{K}_{+}\mathcal{PT}\mathcal{=}\hat{K}_{+},\mathcal{PT}\hat{K}_{-}%
\mathcal{PT}\mathcal{=}\hat{K}_{-}%
\end{equation}
$..$and the time reversal $\mathcal{T}$ change the sign of $i$, we deduce that
$\hat{H}(t)$ is pseudo $\mathcal{PT}$-symmetric, that is,%

\begin{equation}
\mathcal{PT}H(t)\mathcal{PT}\mathcal{=}\Omega(t)\hat{K}_{0}-iG(t)\left(
\hat{K}_{+}+\hat{K}_{-}\right)  =H^{^{\dag}}(t).
\end{equation}
In both time-independent and time-dependent cases, the Hermitian
metric\textbf{ }$\eta(t)=\eta^{+}(t)$\textbf{ }is not uniquely defined. In
order to solve the time Heisenberg equation (\ref{Heisenberg}), I make for
simplicity, the following ansatz for the metric operator $\eta(t)$ such as \ %

\begin{equation}
\eta(t)=\exp\left(  i\gamma(t)\left[  \hat{K}_{+}-\hat{K}_{-}\right]  \right)
,\text{ } \label{1.1}%
\end{equation}
and
\begin{equation}
\eta^{-1}(t)=\exp\left(  -i\gamma(t)\left[  \hat{K}_{+}-\hat{K}_{-}\right]
\right)  , \label{1.2}%
\end{equation}
where $\gamma(t)\ $is a real parameter to be determined.

One can write the operator\textrm{ }$\eta(t)$ in the disentangled form
\cite{kli}%

\begin{equation}
\eta(t)=\exp\left(  v\hat{K}_{+}\right)  \exp\left(  \ln v_{0}\hat{K}%
_{0}\right)  \exp\left(  -v^{\ast}\hat{K}_{-}\right)  \label{dis}%
\end{equation}
where
\begin{equation}
\nu=i\tanh\gamma\ \text{and }v_{0}=(1+\ln\left\vert \nu\right\vert ^{2}),
\label{dis1}%
\end{equation}
and with the help of Baker--Campbell--Hausdorff's formula%
\begin{equation}
\exp(A)B\exp-(A)=B+\left[  A,B\right]  +\frac{1}{2!}\left[  A,\left[
A,B\right]  \right]  +\frac{1}{3!}\left[  A,\left[  A,\left[  A,B\right]
\right]  \right]  +....,
\end{equation}
and the commutation relations Eq.(\ref{commute}), we can easily obtain the
following relations \cite{lai1,Lai,Maamache98}%
\begin{equation}
\left\{
\begin{array}
[c]{c}%
\eta\hat{K}_{+}\eta^{-1}=\hat{K}_{+}\cos%
{{}^2}%
(\gamma)-\hat{K}_{-}\sin%
{{}^2}%
(\gamma)-i\hat{K}_{0}\sin(2\gamma),\\
\eta\hat{K}_{-}\eta^{-1}=\hat{K}_{-}\cos%
{{}^2}%
(\gamma)-\hat{K}_{+}\sin%
{{}^2}%
(\gamma)-i\hat{K}_{0}\sin(2\gamma),\\
\eta\hat{K}_{0}\eta^{-1}=\hat{K}_{0}\cos(2\gamma)-\frac{i}{2}\sin
(2\gamma)\left[  \hat{K}_{+}+\hat{K}_{-}\right]
\end{array}
\right.  \label{1.3}%
\end{equation}

Note that the operators\textbf{ }$\mathcal{\hat{K}}_{0}=$\textbf{ }$\eta
\hat{K}_{0}\eta^{-1}$\textbf{, }$\mathcal{\hat{K}}_{\pm}=\eta\hat{K}_{\pm}%
\eta^{-1}$\textbf{ }in equation Eq.(\ref{1.3}) satisfy an extended
pseudo-commutation relations similar to those in equation (\ref{commute})%
\begin{align}
\left[  \mathcal{\hat{K}}_{0},\mathcal{\hat{K}}_{\pm}\right]   &
=\pm\mathcal{\hat{K}}_{\pm}\,,\nonumber\\
\left[  \mathcal{\hat{K}}_{+},\mathcal{\hat{K}}_{-}\right]   &
=-2\mathcal{\hat{K}}_{0}. \label{commute2}%
\end{align}

The calculation of%
\begin{align}
\eta H(t)\eta^{-1}  &  =\left\{  \Omega\cos(2\gamma)+2G(t)\sin(2\gamma
)\right\}  \hat{K}_{0}\nonumber\\
&  +i\left\{  G\cos(2\gamma)-\frac{\Omega}{2}\sin(2\gamma)\right\}  \hat
{K}_{+}e^{+i\beta}\nonumber\\
&  +i\left\{  G\cos(2\gamma)-\frac{\Omega}{2}\sin(2\gamma)\right\}  \hat
{K}_{-}e^{-i\beta}, \label{6'}%
\end{align}
and%
\begin{equation}
i\frac{\partial\eta}{\partial t}\eta^{-1}=-\dot{\gamma}\left[  \hat{K}%
_{+}-\hat{K}_{-}\right]  , \label{8}%
\end{equation}
gives%

\begin{align}
\eta H(t)\eta^{-1}+i\frac{\partial\eta}{\partial t}\eta^{-1}  &  =\left[
\Omega\cos(2\gamma)+2G(t)\sin(2\gamma)\right]  \hat{K}_{0}\nonumber\\
&  +\left[  iG\cos(2\gamma)-\frac{i\Omega}{2}\sin(2\gamma)-\dot{\gamma
}\right]  \hat{K}_{+}\nonumber\\
&  +\left[  iG\cos(2\gamma)-\frac{i\Omega}{2}\sin(2\gamma)+\dot{\gamma
}\right]  \label{sei}%
\end{align}

By applying $\mathcal{PT}$\ on the left and the right of the two sides of the
equation (\ref{sei}) we obtain, after some algebra, the transformed
Hamiltonian,%
\begin{align}
\widetilde{\eta}(t)H(t)\widetilde{\eta}^{-1}(t)-i\frac{\partial\tilde{\eta
}(t)}{\partial t}\text{ }\tilde{\eta}^{-1}(t)=\hat{K}_{0}\left[  \Omega
\cos(2\gamma)+2G\sin(2\gamma)\right]   &  +\hat{K}_{+}\left[  -iG\cos
(2\gamma)+\frac{i\Omega}{2}\sin(2\gamma)-\dot{\gamma}\right] \\
&  +\hat{K}_{-}\left[  -iG\cos(2\gamma)+\frac{i\Omega}{2}\sin(2\gamma
)+\dot{\gamma}\right]
\end{align}

\begin{align}
\widetilde{\eta}(t)H(t)\widetilde{\eta}^{-1}(t)-i\frac{\partial\tilde{\eta
}(t)}{\partial t}\text{ }\tilde{\eta}^{-1}(t)  &  =\left[  \Omega\cos
(2\gamma)+2G\sin(2\gamma)\right]  \hat{K}_{0}\nonumber\\
&  +\left[  -iG\cos(2\gamma)+\frac{i\Omega}{2}\sin(2\gamma)-\dot{\gamma
}\right]  \hat{K}_{+}\nonumber\\
&  +\left[  -iG\cos(2\gamma)+\frac{i\Omega}{2}\sin(2\gamma)+\dot{\gamma
}\right]  \hat{K}_{-}\nonumber\\
&  =H(t). \label{D'}%
\end{align}

Establishing a connection between the two equations .(\ref{0}) and (\ref{D'})
and through identification, it becomes evident that :%
\begin{equation}
\left\{
\begin{array}
[c]{c}%
\dot{\gamma}=0\\
\Omega(t)=2G(t)\frac{\cos(\gamma)}{\sin(\gamma)}%
\end{array}
\right.  \label{10}%
\end{equation}
where $\gamma$ is a constant. The solution of the Heisenberg equation
(\ref{Heisenberg}),  for the example studied (\ref{0}) as well as the choice
of the metric (\ref{1.1}), leads to a time-independent metric operator $\eta$
and consequently to a time-independent Dyson operator $\rho.$

\subsection{Solution of the Schr\"{o}dinger equation for time-dependent
$SU(1,1)$ Hamiltonian}

\bigskip\ In order to solve the Schr\"{o}dinger equation (\ref{SE})\textrm{,
\ }where the Hamiltonian $\hat{H}(t)$ is obtained in equation (\ref{D'}), we
shall adapt to this case the Dyson operator $\rho$ given by
\begin{equation}
\rho=\exp\left(  i\frac{\gamma}{2}\left[  \hat{K}_{+}-\hat{K}_{-}\right]
\right)  \mathrm{,}\text{ } \label{11}%
\end{equation}
The operator $\rho(t)$ gives rise to the same transformations Eq. (\ref{D'})
where $\gamma$ is replaced by $\gamma/2$ , applying the operato\^{r} $\rho(t)$
directly to the Schrodinger Eq.(\ref{SE}), we have%

\begin{equation}
i\frac{\partial\left\vert \psi^{\prime}(t)\right\rangle }{\partial
t}=H^{\prime}(t)\left\vert \psi^{\prime}(t)\right\rangle \label{SE'}%
\end{equation}
in which%

\begin{equation}
\left\vert \psi^{\prime}(t)\right\rangle =\rho\left\vert \psi(t)\right\rangle
\label{12}%
\end{equation}
It then follows immediately that
\begin{equation}
H^{\prime}(t)=\rho H(t)\rho^{-1} \label{13}%
\end{equation}

We refer to Eq. (\ref{13}) as the time-dependent Dyson relation as it
generalizes its time-independent counterpart \textbf{we} also need an
additional transformation (\ref{8}) where $\gamma$ should be substituted by
$\gamma/2$ . All this yields the Hamitonian \ %

\begin{align}
H^{\prime}  &  =\left[  \Omega\cos(\gamma)+2G\sin(\gamma)\right]  \hat{K}%
_{0}\nonumber\\
&  =\frac{\Omega(t)}{\cos(\gamma)}\hat{K}_{0}. \label{14}%
\end{align}
With all these ingredients we can get the solution of Schrodinger equation
(\ref{SE})\textrm{ }as
\begin{equation}
\left\vert \psi(t)\right\rangle =\rho^{-1}\left\vert \psi^{\prime
}(t)\right\rangle \mathrm{=}\exp\left(  -ik_{n}\int_{0}^{t}\frac
{\Omega(t^{\prime})}{\cos(\gamma)}dt^{\prime}\right)  \left\vert
n\right\rangle \label{sol}%
\end{equation}
where $\left\vert n\right\rangle $ and $k_{n}$ are the eigenstates and
eigenvalue of the operator $\hat{K}_{0}.$

\section{Conclusion}

As an extension of conventional quantum mechanics into the complex domain, the
concept of parity-time ($\mathcal{PT}$) symmetric Hamiltonian has given new
understanding to the behaviors of non-Hermitian systems. A system described by
the non-Hermitian Hamiltonians can also possess real spectra if it is
$\mathcal{PT}$ symmetric.

Another possibility to explain the reality of the spectrum is making use the
pseudo-Hermiticity elation where a non-Hermitian Hamiltonian, $H\neq H^{\dag}%
$, is related to a Hermitian Hamiltonian, $h=\rho H\rho^{-1}$ , by means of
Dyson operator$\rho$. An important object that can be calculated directly from
the Dyson operator is the Hermitian metric operator $\eta=\rho^{+}\rho
$\ \ that link the non-Hermitian Hamiltonian to its conjugate $H^{\dagger
}=\eta H\eta^{-1}.$

Another option to relate $H$ to $H^{\dag}$ is the concept of the pseudo
parity-time (pseudo-$\mathcal{PT}$ ) symmetry $H^{\dag}\mathcal{=PT}%
H\mathcal{PT}$. \ In order to restore the Hamiltonian $H$\ \ based on the last
two relations (pseudo-Hermiticity and pseudo $\mathcal{PT}$-symmetry), we
introduce the $\eta$ pseudo $\mathcal{PT}$-symmetry where the new symmetry is
defind as $\tilde{\eta}=\mathcal{PT}\eta$\ for more details see
Ref.\cite{Absi}.

A many concrete Hamiltonian systems can not be described by autonomous
Hamiltonians $H$, but require an explicit dependence on time $H(t)$. In this
work we have proposed a method for the derivation of general continuous
symmetry operators for a pseudo-$\mathcal{PT}$ symmetric time-dependent
non-Hermitian Hamiltonians. We have used simultaneously pseudo $\mathcal{PT}$
symmetric and $\eta$\ -pseudo-Hermiticity for the time dependent non-Hermitian
Hamiltonian and have defined the relation $\tilde{\eta}(t)=\mathcal{PT}\eta
$\ $(t)$ in order to show that the new symmetry $\tilde{\eta}(t)$ obey the
Heisenberg equation and not the time-dependent quasi-Hermiticity relation
defined in \cite{7', 8-}. Therefore, we applied our theory to the
time-dependent non-Hermitian $SU(1,1)$ system and thus obtained in a simple
way the solutions for Eq. \ref{Heisenberg} of $\tilde{\eta}(t)$ and also the
Schrodinger equation.

I hope that this finding can contribute to any novel effects in experiments in
the futur.

\paragraph{Acknowledgements}

I would like to thank Prs. Philip D. Mannheim and Bryan. W. Roberts for the
explanations provided concerning the time reversal operator in quantum
mechanics and N. El Houda Absi it's valuable discussions.

\end{document}